\documentclass[aps,prb,twocolumn,showpacs,superscriptaddress]{revtex4}
\usepackage{graphicx}
\usepackage{amsmath}
\usepackage{amssymb}
\usepackage{verbatim}

\newcommand{\bm}{\mathbf}
\newcommand{\tm}{\mathrm}

\begin{document}

\title{Momentum dependence of the spin susceptibility in two dimensions: \\ nonanalytic corrections in the Cooper channel}

\author{Stefano Chesi}
\affiliation{Department of Physics, University of Basel, Klingelbergstrasse 82, CH-4056 Basel, Switzerland}
\author{Robert Andrzej \.Zak}
\affiliation{Department of Physics, University of Basel, Klingelbergstrasse 82, CH-4056 Basel, Switzerland}
\author{Pascal Simon}
\affiliation{Department of Physics, University of
Basel, Klingelbergstrasse 82, CH-4056 Basel, Switzerland}
\affiliation{Laboratoire de Physique et Mod\'elisation des Milieux Condens\'es, CNRS, Universit\'e Joseph Fourier, BP 166, 38042 Grenoble, France}
\affiliation{Laboratoire de Physique des Solides, CNRS UMR-8502 Universit\'e Paris Sud, 91405 Orsay Cedex, France}
\author{Daniel Loss}
\affiliation{Department of Physics, University of
Basel, Klingelbergstrasse 82, CH-4056 Basel, Switzerland}

\begin{abstract}
We consider the effect of rescattering of pairs of quasiparticles in the Cooper channel resulting in the strong renormalization of second-order corrections to the spin susceptibility in a two-dimensional electron system. We use the Fourier expansion of the scattering potential in the vicinity of the Fermi surface to find that each harmonic becomes renormalized independently. Since some of those harmonics are negative, the first derivative of the spin susceptibility is bound to be negative at small momenta, in contrast to the lowest order perturbation theory result, which predicts a positive slope. We present in detail an effective method to calculate diagrammatically corrections to the spin susceptibility to infinite order.
\end{abstract}


\maketitle

%%%%%%%%%%%%%%%%%%%%%%%%%%%%%%%%%%%%%%%%%%%%%%%%%%%%%%%%%%%%%%%%%%%%%%%%%%%%%%%%%%%%%%%
\section{\label{sec:int}INTRODUCTION}
%%%%%%%%%%%%%%%%%%%%%%%%%%%%%%%%%%%%%%%%%%%%%%%%%%%%%%%%%%%%%%%%%%%%%%%%%%%%%%%%%%%%%%%

The study of the thermodynamic as well as microscopic properties of Fermi-liquid systems has a long history,\cite{landau57,landau59,pines,gfg} but the interest in nonanalytic corrections to the Fermi-liquid behavior is more recent. The existence of well-defined quasiparticles at the Fermi surface is the basis for the phenomenological description due to Landau\cite{landau57} and justifies the fact that a system of interacting fermions is similar in many ways to the Fermi gas. The Landau theory of the Fermi liquid is a fundamental paradigm which has been successful in describing properties of ${}^3$He, metals, and two-dimensional electronic systems. In particular, the leading temperature dependence of the specific heat or the spin susceptibility (i.e., $C_s$ linear in $T$ and $\chi_s$ approaching a constant) is found to be valid experimentally and in microscopic calculations. However, deviations from the ideal Fermi gas behavior exist in the subleading terms.

For example, while the low-temperature dependence of $C_s/T$ for a Fermi gas is a regular expansion in $T^2$, a correction to $C_s/T$ of the form $T^2 \ln T$ was found in three dimensions.\cite{pethick73} These nonanalytic features are enhanced in two dimensions and, in fact, a correction linear in $T$ is found.\cite{coffey93,PhysRevB.55.9452,PhysRevB.68.155113} These effects were observed in ${}^3$He, both in the three-\cite{greywall83} and two-dimensional case.\cite{casey03}

The nonanalytic corrections manifest themselves not only in the temperature dependence. For the special case of the spin susceptibility, it is of particular interest to determine also its dependence on the wave vector $Q$. The deviation $\delta\chi_s$ from the $T=Q=0$ value parallels the temperature dependence of the specific heat discussed above: from a second-order calculation in the electron interaction, corrections proportional to $Q^2\ln Q$ and $Q$ were obtained in three and two dimensions respectively.\cite{PhysRevB.55.9452,hirashima98,PhysRevB.68.155113} On the other hand, the dependence on $T$ was found to be $\delta\chi_s \sim T^2$ in three dimensions\cite{PhysRevB.16.1933,PhysRevB.55.9452} (without any logarithmic factor) and $\delta\chi_s \sim T$ in two dimensions.\cite{hirashima98,JETPLett.58.709,PhysRevLett.86.5337,PhysRevB.64.054414,PhysRevB.68.155113}
We cite here the final results in the two dimensional case (on which we focus in this paper), valid to second order in the interaction potential $V(q)$,
\begin{equation}\label{eq:deltachi_2}
    \delta\chi_s^{(2)}(T,Q)=2K(T,Q)V^{2}(2 k_F),
\end{equation}
where
\begin{equation}\label{eq:KT}
    K(T,0)=\frac{m^3}{16\pi^3}\frac{k_B T}{E_{F}}
\end{equation}
and
\begin{equation}\label{eq:KQ}
    K(0,Q)\equiv \frac{m^3}{48\pi^4}\frac{v_F Q}{E_{F}}.
\end{equation}
Here $m$ is the effective mass, $k_F$ is the Fermi wave vector, $E_F=k_F^2/2m$, and we use $\hbar=1$ throughout the paper. Our purpose is to extend this perturbative result to higher order by taking into account the Cooper channel renormalization of the scattering amplitudes. 

The extension to higher order of the second-order results has mostly focused on the temperature dependence, both for the specific heat\cite{chubukov05a,chubukov05b,chubukov06,chubukov07,aleiner06} and the spin susceptibility.\cite{chubukov05a,PhysRevB.74.205122,ProcNatlAcadSci.103.15765,schwiete06} Recently the spin susceptibility has been measured in a silicon inversion layer as a function of temperature.\cite{PhysRevB.67.205407} A strong dependence on $T$ is observed, seemingly incompatible with a $T^2$ Fermi-liquid correction, and the measurements also reveal that the (positive) value of the spin susceptibility is \emph{decreasing} with temperature, in disagreement with the lowest order result cited above. This discrepancy has stimulated further theoretical investigations in the nonperturbative regime. Possible mechanisms that lead to a negative slope were proposed in Refs.~\onlinecite{PhysRevB.74.205122} and \onlinecite{ProcNatlAcadSci.103.15765} if strong renormalization effects in the Cooper channel become important. These can drastically change the picture given by the lowest order perturbation theory, allowing for a nonmonotonic behavior and, in particular, a negative slope at small temperatures. 

The mechanism we consider here to modify the linear $Q$ dependence is very much related to Ref.~\onlinecite{PhysRevB.74.205122}. There it is found that, at $Q=0$ and finite temperature, $V^2(2k_F)$ in Eq.~(\ref{eq:deltachi_2}) is substituted by $|\Gamma(\pi)|^{2}$, where 
\begin{equation}\label{eq:Gamma_theta_def}
    \Gamma(\theta)\equiv\sum_{n}\Gamma_{n}e^{in\theta}
\end{equation}
is the scattering amplitude in the Cooper channel with $\theta$ being the scattering angle ($\theta=\pi$ corresponds to the backscattering process). An additional temperature dependence arises from the renormalization of the Fourier amplitudes
\begin{equation}\label{eq:GammaT}
    \Gamma_{n}(k_{B}T) = \frac{V_{n}}{1-\frac{m V_{n}}{2\pi}\ln\frac{k_B T}{W}},
\end{equation}
where $W$ is a large energy scale $W\sim E_{F}$ and $V_n$ are the Fourier amplitudes of the interaction potential for scattering in the vicinity of the Fermi surface
\begin{equation}\label{eq:Vdef}
    V(2k_{F}\sin{\theta/2}) = \sum_{n}V_{n}e^{in\theta}.
\end{equation}
A negative slope of $\delta\chi_s$ is possible, for sufficiently small $T$ if one of the amplitudes $V_n$ is negative.\cite{PhysRevB.74.205122,PhysRevLett.15.524,PhysRevB.48.1097} For $\frac{m V_{n}}{2\pi}\ln\frac{k_B T_{KL}}{W}=1$, the denominator in Eq.~(\ref{eq:GammaT}) diverges what corresponds to the Kohn-Luttinger (KL) instability.\cite{PhysRevLett.15.524} At $T\gtrsim T_{KL}$ the derivative of the spin susceptibility is negative due to the singularity in $\Gamma_{n}(k_{B}T)$ and becomes positive far away from $T_{KL}$.
%There are general reasons why the negative amplitudes are always present for some harmonics.\cite{PhysRevLett.15.524,PhysRevB.48.1097

At $T=0$ an analogous effect occurs for the momentum dependence. Indeed, it is widely expected that the functional form of the spin susceptibility in terms of $k_B T$ or $v_F Q$ is similar. As in the case of a finite temperature, the lowest order expression gains an additional nontrivial dependence on $Q$ due to the renormalization of the backscattering amplitude $V^2(2k_F)$. We obtain
\begin{equation}\label{eq:MainRes}
    \delta\chi_s(Q)=2K(0,Q)|\Gamma(\pi)|^2,
\end{equation}
where $\Gamma(\pi)$ is given by Eq.~(\ref{eq:Gamma_theta_def}) and
\begin{equation}\label{eq:GammaDef}
\Gamma_n(v_{F}Q)=\frac{V_n}{1-\frac{m V_n}{2\pi}\ln\frac{v_F Q}{W}}.
\end{equation}
Such result is obtained from renormalization of the interaction in the Cooper channel, while other possible effects are neglected. Moreover, at each perturbative order, only the leading term in the limit of small $Q$ is kept. Therefore, corrections to  Eq.~(\ref{eq:MainRes}) exist which, for example, would modify the proportionality of $\delta\chi_s$ to $|\Gamma(\pi)|^2$ (see Ref.~\onlinecite{PhysRevB.74.205122}). However, in the region $v_F Q \gtrsim k_B T_{KL}$, close to the divergence of $\Gamma_n(v_{F}Q)$ relative to the most negative $V_n$, Eq.~(\ref{eq:MainRes}) is expected to give the most important contribution to the spin susceptibility.

The result of Eqs.~(\ref{eq:MainRes}) and (\ref{eq:GammaDef}) could have been perhaps easily anticipated and, in fact, it was suggested already in Ref.~\onlinecite{PhysRevB.77.045108}. The question of the functional dependence of the spin susceptibility on momentum is crucial in light of the ongoing studies on the nuclear spin ferromagnetism,\cite{PhysRevLett.98.156401,PhysRevB.77.045108,PhysRevB.67.144520} as the stability of the ferromagnetic phase is governed by the electron spin susceptibility. In this context, Eqs.~(\ref{eq:MainRes}) and (\ref{eq:GammaDef}) were motivated by a renormalization-group argument. We provide here a complete derivation, based on the standard diagrammatic approach.
 
The paper is organized as follows: in Sec. \ref{sec:ppp} we discuss the origin of Cooper instability and derive expressions for a general ladder diagram, which is an essential ingredient for the higher order corrections to the spin susceptibility. In Sec.~\ref{sec:2od} we give a short overview of the lowest order results to understand the origin of the nonanalytic corrections. Based on the results of Section \ref{sec:ppp}, we provide an alternative derivation of one of the contributions, which can be easily generalized to higher order. 
Sec.~\ref{sec:hod} contains the main finding of this paper: the Cooper renormalization of the nonanalytic correction to the spin susceptibility is obtained there. We find an efficient approach to calculate higher order diagrams based on the second-order result. In Section \ref{sec:RG} the diagrammatic calculation is discussed in relation to the renormalization-group 
argument of Ref.~\onlinecite{PhysRevB.77.045108}. Sec.~\ref{sec:con} contains our concluding remarks. More technical details have been moved to the Appendixes~\ref{app:lad}-\ref{app:ppsmallQ}.

%%%%%%%%%%%%%%%%%%%%%%%%%%%%%%%%%%%%%%%%%%%%%%%%%%%%%%%%%%%%%%%%%%%%%%%%%%%%%%%%%%%%%%
\section{\label{sec:ppp}PARTICLE-PARTICLE PROPAGATOR}
%%%%%%%%%%%%%%%%%%%%%%%%%%%%%%%%%%%%%%%%%%%%%%%%%%%%%%%%%%%%%%%%%%%%%%%%%%%%%%%%%%%%%%

\begin{figure}
    \includegraphics[width=.4\textwidth]{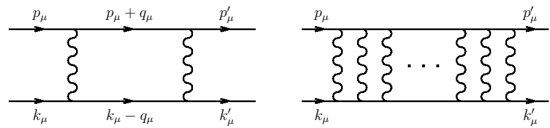}
    \caption{\label{fig:2}The building block (on the left) of any ladder diagram (on the right). Of special interest
    is the limit of correlated momenta $\bm{p}=-\bm{k}$, leading to the Cooper instability.}
\end{figure}

In this section we consider a generic particle-particle propagator, which includes $n$ interaction lines, as depicted in Fig.~\ref{fig:2}. The incoming and outgoing frequencies and momenta are $k_\mu, p_\mu$ and $k'_\mu, p'_\mu$, respectively, using the relativistic notation $k_\mu=(\omega_{k},\bm{k})$. This particle-particle propagator represents an essential part of the diagrams considered in this paper and corresponds to the following expression:
\begin{align}\label{eq:Pidef}
    \notag \Pi&^{(n)}(p_{\mu},p_{\mu}',k_{\mu}) = (-1)^{n-1}\int\frac{d^{3} q_{1} \dots
   d^{3}q_{n-1}}{(2\pi)^{3n-3}} V(|{\bf q}_1|) \\
    &\times \prod_{i=1}^{n-1}G(k_{\mu}-q_{i,\mu})G(p_{\mu}+q_{i,\mu})V(|\bm{q}_{i+1}-\bm{q}_{i}|),
\end{align}
where $\bm{q}_{n}\equiv\bm{p}'-\bm{p}$. The frequencies are along the imaginary axis, i.e., $G(k_\mu)=G(\omega_{k},\bm{k})=(i\omega_{k}-\xi_{\bm{k}})^{-1}$, where $\xi_{\bm{k}}=k^2/2m-E_F$ with $k=|{\bf k}|$.   

%%%%%%%%%%%%%%%%%%%%%%%%%%%%%%%%%%%%%%%%%%%%%%%%%%%%%%%%%%%%%%%%%%%%%%%%%%%%%%%%%%%%%%
%%%%%%%%%%%%%%%%%%%%%%%%%%%%%%%%%%%%%%%%%%%%%%%%%%%%%%%%%%%%%%%%%%%%%%%%%%%%%%%%%%%%%%

In particular, we are interested in the case when the sum of incoming frequencies and momenta is small; i.e., $P_\mu\equiv p_\mu+k_\mu \approx0$. Under this assumption we obtain the following useful result for which we provide details of the derivation in Appendix~\ref{app:lad}:
\begin{align}\label{eq:PiNres}
    \notag \Pi^{(n)}(P_\mu,\theta) = {\sum_{M_{1}\dots M_{n-1}}}^{\hspace{-10pt}\prime}&
    ~\Pi_{M_{1}}(P_\mu)\dots\Pi_{M_{n-1}}(P_\mu)\\
    &\times\tilde{V}_{M_{1}\dots M_{n-1}}^{n}(\theta_P, \theta),
\end{align}
where the sum is restricted to $M_i=0,\pm2,\pm4 \ldots$. The angle of ${\bf P}={\bf p}+\bf{k}$ is from the direction of the incoming momentum ${\bf p}$,  i.e., $\theta_{P}\equiv\angle(\bm{P},\bm{p})$, while $\theta\equiv\angle(\bm{p}',\bm{p})$. In the above formula,
\begin{equation}\label{eq:Pi0}
    \Pi_{0}(P_\mu) = \frac{m}{2\pi}
    \ln{\frac{|\Omega_{P}|+\sqrt{\Omega_{P}^{2}+v_{F}^{2}P^{2}}}{W}}\qquad
\end{equation}
and ($M$ even)
\begin{equation}\label{eq:PiM}
    \Pi_{M\neq0}(P_\mu) = -\frac{m}{2\pi}\frac{(-1)^{|M|/2}}{|M|}
    \Big(\frac{1-\sin\phi}{\cos\phi}\Big)^{|M|},
\end{equation}
with $W \sim E_F$ a high energy cutoff and $\phi\equiv\arctan\frac{|\Omega_{P}|}{v_{F}P}$. Notice that $\Pi_M(P_\mu)$ has no angular ($\theta_P$, $\theta$) dependence, which is only determined by the following quantity:
\begin{align}\label{eq:VMNdef}
    \notag \tilde{V}_{M_{1}\dots M_{n-1}}^{n}&(\theta_P,\theta)
    \equiv \sum_{m,m'}V_{m}V_{m-M_{1}}\dots V_{m-M_{1}-\ldots M_{n-1}}\\
    &\times e^{i m'\theta_{P}- i \, m \theta} \, \delta_{M_{1}+M_2+\ldots M_{n-1},m'}
\end{align}
defined in terms of the amplitudes $V_n$. Equation~(\ref{eq:Vdef})
can be used to approximate the interaction potential in Eq.~(\ref{eq:Pidef}) since the relevant contribution originates from the region of external ($p \approx p'\approx k \approx k' \approx k_{F}$) and
internal momenta ($|\bm{p}+\bm{q}_i|\approx|\bm{k}-\bm{q}_i|\approx k_{F}$) close to the Fermi surface.
Furthermore, the direction of ${\bf P}$ can be equivalently measured from $\bf{k}$ without affecting the result since $\theta_{P}=\angle(\bm{P},\bm{k})+\pi$ and $e^{im'\pi}=1$ ($m'$ is even).

Notice also that the leading contribution to Eq.~(\ref{eq:PiNres}), in the limit of small $\Omega_P$ and $P$, is determined by the standard logarithmic singularity of $\Pi_{0}(P_\mu)$. However, it will become apparent that this leading contribution is not sufficient to obtain the correct result for the desired (linear-in-$Q$) corrections to the response function. The remaining terms, $\Pi_{M}(P_\mu)$, are important because of their nonanalytic form due to the dependence on the ratio $\frac{|\Omega_{P}|}{v_{F}P}$.

%%%%%%%%%%%%%%%%%%%%%%%%%%%%%%%%%%%%%%%%%%%%%%%%%%%%%%%%%%%%%%%%%%%%%%%%%%%%%%%%%%%%%%%
\section{\label{sec:2od} SECOND-ORDER CALCULATION}
%%%%%%%%%%%%%%%%%%%%%%%%%%%%%%%%%%%%%%%%%%%%%%%%%%%%%%%%%%%%%%%%%%%%%%%%%%%%%%%%%%%%%%%

\begin{figure}
    \includegraphics[width=.4\textwidth]{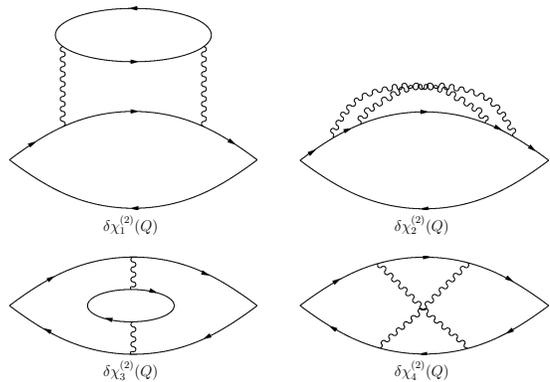}
    \caption{\label{fig:1}The nonvanishing second-order diagrams contributing to the nonanalytic behavior
    of the electron spin susceptibility.}
\end{figure}

The lowest order nonanalytic correction to the spin susceptibility has been calculated in Ref.~\onlinecite{PhysRevB.68.155113} as a sum of four distinct contributions from the diagrams in Fig.~\ref{fig:1},
\begin{align}
    \delta\chi^{(2)}_{1}(Q)&=K(0,Q)[V^{2}(2k_F)+V^{2}(0)]\label{eq:chi21},\\
    \delta\chi^{(2)}_{3}(Q)&=K(0,Q)[V^{2}(2k_F)-V^{2}(0)]\label{eq:chi23},\\
    \delta\chi^{(2)}_{4}(Q)&=K(0,Q)V(0)V(2 k_F)\label{eq:chi24},
\end{align}
and $\delta\chi^{(2)}_{2}=-\delta\chi^{(2)}_{4}$ such that the final result reads as
\begin{equation}\label{eq:2nd_order_final}
    \delta\chi_s^{(2)}(Q)=2K(0,Q)V^{2}(2k_F).
\end{equation}

%%%%%%%%%%%%%%%%%%%%%%%%%%%%%%%%%%%%%%%%%%%%%%%%%%%%%%%%%%%%%%%%%%%%%%%%%%%%%%%%%%%%%%%%%%%
%%%%%%%%%%%%%%%%%%%%%%%%%%%%%%%%%%%%%%%%%%%%%%%%%%%%%%%%%%%%%%%%%%%%%%%%%%%%%%%%%%%%%%%%%%%

We refer to Ref.~\onlinecite{PhysRevB.68.155113} for a thorough discussion of these lowest order results, but we find it useful to reproduce here the result for $\delta\chi_1^{(2)}$. In fact, Eq.~(\ref{eq:chi21}) has been obtained in Ref.~\onlinecite{PhysRevB.68.155113} as a sum of two nonanalytic contributions from the particle-hole bubble at small ($q=0$) and large ($q=2k_{F}$) momentum transfer. These two contributions, proportional to $V^{2}(0)$ and $V^{2}(2k_F)$, respectively, can be directly seen in Eq.~(\ref{eq:chi21}). However, it is more natural for our purposes to obtain the same result in the particle-particle channel by making use of the propagator discussed in Sec.~\ref{sec:ppp}. This approach is more cumbersome but produces these two contributions at the same time. Furthermore, once the origin of the lowest order nonanalytic correction is understood in the particle-particle channel, higher order results are most easily obtained.

%%%%%%%%%%%%%%%%%%%%%%%%%%%%%%%%%%%%%%%%%%%%%%%%%%%%%%%%%%%%%%%%%%%%%%%%%%%%%%%%%%%%%%%%%%%
\begin{figure}
    \includegraphics[width=.2\textwidth]{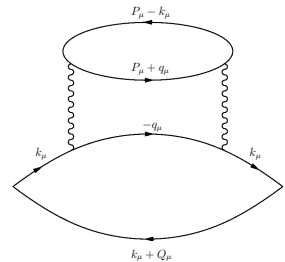}
    \caption{\label{fig:0} Labeling of the $\delta\chi_1^{(2)}$ diagram, as in Eq.~(\ref{eq:chi1-2def}).}
\end{figure}
%%%%%%%%%%%%%%%%%%%%%%%%%%%%%%%%%%%%%%%%%%%%%%%%%%%%%%%%%%%%%%%%%%%%%%%%%%%%%%%%%%%%%%%%%%%

We start with the analytic expression of $\delta\chi_{1}^{(2)}(Q)$ (see Fig.~\ref{fig:0}) in terms of $\Pi^{(2)}$, the $n=2$ case of Eq.~(\ref{eq:PiNres});
\begin{align}\label{eq:chi1-2def}
    \notag\delta\chi_{1}^{(2)}(Q) =& -8\int\frac{d^{3}k}{(2\pi)^{3}}
    \int\frac{d^{3}P}{(2\pi)^{3}}G^{2}(k_{\mu}) G(k_{\mu}+Q_{\mu})\\
    &\times G(P_{\mu}-k_{\mu})\Pi^{(2)}(P_{\mu},0).
\end{align}
It is convenient to define the angle of ${\bf k}$ as $\theta_k\equiv\angle(\bm{k},\bm{Q})$, and $\theta_P\equiv\angle(\bm{P},\bm{k})$. We first perform the integration in $d^3 k$, as explained in Appendix~\ref{app:chi1-2}, to obtain
\begin{align}\label{eq:2ndorder_dimensional}
    \notag &\delta\chi_{1}^{(2)} = -\frac{m}{\pi^{4}v_{F}^{2}Q^{2}}\int_{0}^{\infty}P{ d}P
    \int_{0}^{\infty}{ d}\Omega_{P}\int_{0}^{2\pi} \Pi^{(2)}(P_{\mu},0)\\
    &\times
    \left(1-\frac{\sqrt{(\Omega_{P}+iv_{F}P\cos\theta_P)^{2}+(v_{F}Q)^{2}}}
    {\Omega_{P}+iv_{F}P\cos\theta_P}\right) { d}\theta_P  .
\end{align}
Following the method of Ref.~\onlinecite{PhysRevB.68.155113}, we rescale the integration variables: 
$\Omega_{P}=Rv_{F}Q\sin\phi$, $P=RQ\cos\phi$, and $d\Omega_{P}dP=Rv_{F}Q^{2}dRd\phi$. This gives
\begin{align}\label{eq:chi1-2rescaled}
    \notag & \delta\chi_{1}^{(2)}
    = -\frac{m Q}{\pi^{4}v_{F}}\int_{0}^{\infty}R^{2}{ d}R
    \int_{0}^{\pi/2}{ d}\phi\int_{0}^{2\pi} \Pi^{(2)}(R,\phi,\theta_P,0) \\
    &  \times \cos\phi  \left(1-\frac{\sqrt{R^{2}(\sin\phi+i\cos\phi\cos\theta_P)^{2}+1}}
    {R(\sin\phi+i\cos\phi\cos\theta_P)}\right) { d}\theta_P .
\end{align}
where, from Eqs.~(\ref{eq:PiNres}) and (\ref{eq:VMNdef}), 
\begin{align}\label{eq:ppprop}
    \notag \Pi^{(2)}(R,& \phi,\theta_P, \theta)= 
    {\sum_{M}}^{\prime}\tilde{V}_{M}^{2}(\theta_P,\theta)\Pi_{M}(R,\phi)\\
    = &  {\sum_M}^\prime \Pi_M(R,\phi){\sum_{m}} V_{m}V_{m-M} e^{iM\theta_P-i m \theta}  
\end{align}
with the primed sum restricted to even values of $M$. 

Now we can see clearly that the linear dependence on $Q$ in Eq.~(\ref{eq:chi1-2rescaled}) can only be modified by the presence of $\Pi^{(2)}$ in the integrand because of
\begin{equation}\label{eq:Pi0Rphi}
    \Pi_{0}(R,\phi)=\frac{m}{2\pi}\ln\frac{v_{F}Q}{W}+\frac{m}{2\pi}\ln R(1+\sin\phi).
\end{equation}
The first logarithmic term is diverging at small $Q$ but does not contribute to the final result since it does not depend on $\theta_P$ and $\phi$. In fact, if we keep only the $\frac{m}{2\pi}\ln\frac{v_{F}Q}{W}$ contribution,
after the change of variable $r=R(\sin\phi+i\cos\phi\cos\theta_P)$ in Eq.~(\ref{eq:chi1-2rescaled}), we obtain the angular integral $\int_0^{2\pi}{ d}\theta_P\int_0^{\pi/2}\cos\phi (\sin\phi+i\cos\phi\cos\theta_P)^{-3}  \, { d} \phi=0$ [cf. Eq.~(\ref{eq:intphitheta}) for $M=0$]. Details of the calculation are provided in Appendix~\ref{app:chi1_rederived}.

Therefore, only the second term of Eq.~(\ref{eq:Pi0Rphi}) is relevant. The integral in Eq.~(\ref{eq:chi1-2rescaled}) becomes independent of $Q$ and gives only a numerical prefactor. The final result is given by Eq.~(\ref{eq:chi21}), in agreement with Ref.~\onlinecite{PhysRevB.68.155113}. In a similar way, the remaining diagrams of Fig.~\ref{fig:1} can be calculated.

%%%%%%%%%%%%%%%%%%%%%%%%%%%%%%%%%%%%%%%%%%%%%%%%%%%%%%%%%%%%%%%%%%%%%%%%%%%%%%%%%%%%%%%
\section{\label{sec:hod}HIGHER ORDER DIAGRAMS}
%%%%%%%%%%%%%%%%%%%%%%%%%%%%%%%%%%%%%%%%%%%%%%%%%%%%%%%%%%%%%%%%%%%%%%%%%%%%%%%%%%%%%%%

In this section  we aim to find the renormalization of the four diagrams depicted in Fig.~\ref{fig:1} due to higher order contributions in the particle-particle channel. It is well known that the scattering of two electrons with opposite momenta, in the presence of the Fermi sea, leads to the emergence of a logarithmic singularity.\cite{PhysRevB.71.045338,Mahan} Furthermore, in two dimensions there are just two processes that contribute to $\delta\chi_{i}^{(2)}(Q)$, namely, forward- (small momentum transfer, $q=0$) and back-scattering (large momentum transfer, $q=2k_{F}$). This results in the renormalization of the scattering amplitudes appearing in the second-order results (see Sec.~\ref{sec:int}).

A direct calculation of the particle-particle propagators, depicted in Fig.~\ref{fig:2}, shows that for $n+1$ interaction lines, the divergence always appears as the $n$th power of a logarithm. At each order of the perturbative expansion, we only consider the single diagram which contributes to the nonanalytic correction with the leading logarithmic singularity. This requirement restricts the freedom of adding interaction lines in unfettered manner to the existing second-order diagrams: in order to produce the most divergent logarithmic term, all interaction lines have to build up at most one ladder for $\delta\chi_{1}$, $\delta\chi_{2}$, and $\delta\chi_{4}$, or two ladders for $\delta\chi_{3}$.

The subset of diagrams generated in this way is not sufficient to obtain the general momentum dependence of the spin susceptibility. However, if one of the harmonics $V_{n}$ is negative, these diagrams are the only relevant ones in the vicinity of the Kohn-Luttinger instability, $v_{F}Q\gtrsim k_{B}T_{KL}$. Furthermore, at each order $n$ in the interaction, it suffices to keep the leading contribution in $Q$ of the individual diagrams. This turns out to be of order $Q\ln^{n-2}Q$ because the term proportional to $\ln^{n-1}Q$ is suppressed by an additional factor $Q^2$. Other perturbative terms, e.g., in the particle-hole channel,\cite{ProcNatlAcadSci.103.15765} can be safely neglected as they result in logarithmic factors of lower order. 

In the following we discuss explicitly how to insert a ladder diagram into the pre-existing second-order diagrams and show the line of the calculation that has to be carried out.

%%%%%%%%%%%%%%%%%%%%%%%%%%%%%%%%%%%%%%%%%%%%%%%%%%%%%%%%%%%%%%%%%%%%%%%%%%%%%%%%%%%%%%
\subsection{\label{sec:d124}Diagrams 1, 2, and 4}
%%%%%%%%%%%%%%%%%%%%%%%%%%%%%%%%%%%%%%%%%%%%%%%%%%%%%%%%%%%%%%%%%%%%%%%%%%%%%%%%%%%%%%

\begin{figure}
    \includegraphics[width=.4\textwidth]{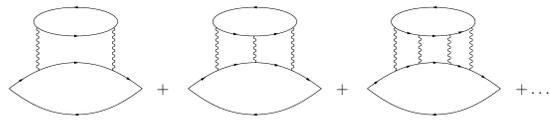}
    \caption{\label{fig:3}The series of diagrams contributing to $\delta\chi_{1}(Q)$.}
\end{figure}
\begin{figure}
    \includegraphics[width=.4\textwidth]{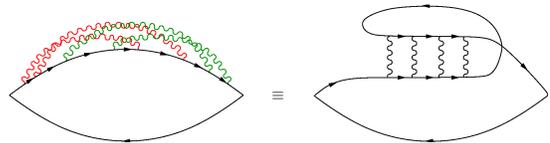}
    \caption{\label{fig:4}An example of diagram contributing to $\delta\chi_{2}(Q)$.
    The maximally crossed diagram (left) is topologically equivalent to its untwisted counterpart (right) in which 			the particle-particle ladder appears explicitly.}
\end{figure}
\begin{figure}
    \includegraphics[width=.4\textwidth]{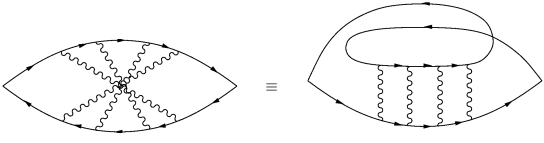}
    \caption{\label{fig:6}A maximally crossed diagram (left) and its untwisted equivalent (right) contributing to
    $\delta\chi_{4}(Q)$.}
\end{figure}

These three diagrams can all be expressed to lowest order in terms of a single particle-particle propagator $\Pi^{(2)}$, which at higher order is substituted by $\Pi^{(n)}$. For the first term we have 
\begin{align}\label{eq:chi1-ndef}
    \notag  \delta\chi_{1}^{(n)}(Q) =& -8\int\frac{{ d}^{3}k}{(2\pi)^{3}}
    \int\frac{{ d}^{3}P}{(2\pi)^{3}}G^{2}(k_{\mu})\\
    &\times G(k_{\mu}+Q_{\mu})G(P_{\mu}-k_{\mu})\Pi^{(n)}(P_{\mu},0),
\end{align}
where the $n=2$ case was calculated in Sec.~\ref{sec:2od}. The corresponding diagrams are, in this case, easily identified and shown in Fig.~\ref{fig:3}. 

It is slightly more complicated to renormalize $\delta\chi_{2}^{(2)}$ and $\delta\chi_{4}^{(2)}$. It requires one to realize that the diagrams depicted in Fig.~\ref{fig:4} are topologically equivalent; i.e., the maximally crossed diagram on the left is equivalent to the untwisted ladder diagram on the right. A similar analysis shows how to lodge the ladder diagram into $\delta\chi_{4}^{(2)}$, as illustrated in Fig.~\ref{fig:6}. The corresponding analytic expressions are:
\begin{align}\label{eq:chi2-ndef}
    \notag  \delta\chi_{2}^{(n)}(Q) & = 4\int\frac{{ d}^{3}k}{(2\pi)^{3}}
    \int\frac{{ d}^{3}P}{(2\pi)^{3}}G^{2}(k_{\mu})\\
     &\times G(k_{\mu}+Q_{\mu}) G(P_{\mu}-k_{\mu})\Pi^{(n)}(P_{\mu},\pi),
\end{align}
\begin{align}\label{eq:chi4-ndef}
    \notag  & \delta \chi_{4}^{(n)}(Q) = 2\int\frac{{ d}^{3}k}{(2\pi)^{3}}
    \int\frac{{ d}^{3}P}{(2\pi)^{3}}G(k_{\mu})G(k_{\mu}+Q_{\mu})\\
    &\times G(P_{\mu}-k_{\mu})G(P_\mu-k_{\mu}-Q_{\mu})\Pi^{(n)}(P_{\mu},\pi).
\end{align}

We show now that the final results can be simply obtained to leading order in $Q$ based on the second-order calculation. In fact, we can perform the integration in ${ d}^3 k$ and the rescaling of variables as before. For $\delta\chi_{1}$ we have
\begin{align}
    \notag &\delta\chi_{1}^{(n)}
    = -\frac{m Q}{\pi^{4}v_{F}}\int_{0}^{\infty}R^{2}{ d}R
    \int_{0}^{\pi}{d}\theta_P\int_{0}^{\pi/2} \Pi^{(n)}(R,\phi,\theta_P,0)\\
    &\times\bigg(1-\frac{\sqrt{R^{2}(\sin\phi+i\cos\phi\cos\theta_P)^{2}+1}}
    {R(\sin\phi+i\cos\phi\cos\theta_P)}\bigg)\cos\phi ~ d\phi .
\end{align}
In the above formula, the $Q$ dependence in the integrand is only due to $\Pi^{(n)}$. It is clear that a similar situation occurs for the second and fourth diagrams.

The $Q$ dependence of the rescaled Eq.~(\ref{eq:PiNres}) is determined (as in the second order) by the factors $\Pi_0(R,\phi)$. The first term appearing in $\Pi_0(R,\phi)$, see Eq.~(\ref{eq:Pi0Rphi}), is large in the small $Q$ limit we are interested in. Therefore, we can expand $\Pi^{(n)}$ in powers of $\frac{m}{2\pi}\ln\frac{v_{F}Q}{W}$ and retain at each perturbative order $n$ only the most divergent nonvanishing contribution. The detailed procedure is explained in Appendix~\ref{app:ppsmallQ}. It is found that the largest contribution from $\Pi^{(n)}$ is of order $(\frac{m}{2\pi}\ln\frac{v_{F}Q}{W})^{n-1}$. However, as in the case of the second-order diagram discussed in Sec.~\ref{sec:2od}, this leading term has an analytic dependence on $P_\mu$ (in fact, it is a constant), and gives a vanishing contribution to the linear-in-$Q$ correction to the spin susceptibility. Therefore, the $(\frac{m}{2\pi}\ln\frac{v_{F}Q}{W})^{n-2}$ contribution is relevant here.

A particularly useful expression is obtained upon summation of $\Pi^{(n)}$ to infinite order. In fact, for each diagram, the sum of the relative series involves the particle-particle propagator only. Therefore, $\delta\chi_{1}$, $\delta\chi_{2}$, and $\delta\chi_{4}$ are given by Eqs.~(\ref{eq:chi1-ndef})--(\ref{eq:chi4-ndef}) if $\Pi^{(n)}$ is substituted by
\begin{equation}
\Pi^{(\infty)}(P_{\mu},\theta)=\sum_{n=2}^\infty \Pi^{(n)}(P_{\mu},\theta).
\end{equation} 
The relevant contribution of $\Pi^{(\infty)}(P_{\mu},\theta)$, in the rescaled variables, is derived in Appendix~\ref{app:ppsmallQ}. The final result is
\begin{align}\label{eq:finalPinSum}
\notag & \Pi^{(\infty)} (R,\phi,\theta_P,\theta)=\sum_{n=2}^\infty \Pi^{(n)} (R,\phi,\theta_P,\theta)\\
 = &  {\sum_M}^\prime \Pi_M(R,\phi)
{\sum_{m}} \Gamma_{m}\Gamma_{m-M} e^{iM\theta_P-im\theta}+\ldots,
\end{align}
which should be compared directly to Eq.~(\ref{eq:ppprop}). The only difference is the replacement of $V_n$ with the renormalized amplitudes $\Gamma_n$, which depend on $Q$ as in Eq.~(\ref{eq:GammaDef}).

Hence, it is clear that the final results follow immediately from Eqs.~(\ref{eq:chi21})--(\ref{eq:chi24});
\begin{align}
    \delta\chi_{1}(Q)&=K(0,Q)[\Gamma^{2}(0)+\Gamma^{2}(\pi)]\label{eq:chi1},\\
    \delta\chi_{4}(Q)&=K(0,Q)\Gamma(0)\Gamma(\pi)\label{eq:chi2and4},
\end{align}
and $\delta\chi_{2}(Q)=-\delta\chi_{4}(Q)$. We have used notation (\ref{eq:Gamma_theta_def}) while $K(0,Q)$ is defined in Eq.~(\ref{eq:KQ}). This explicitly proves what was anticipated in Sec~\ref{sec:int} (and in Ref.~\onlinecite{PhysRevB.77.045108}), i.e., that the renormalization affects only the scattering amplitude. The bare interaction potential is substituted by the dressed one, which incorporates the effect of other electrons on the scattering pair.

%%%%%%%%%%%%%%%%%%%%%%%%%%%%%%%%%%%%%%%%%%%%%%%%%%%%%%%%%%%%%%%%%%%%%%%%%%%%%%%%%%%%%%
\subsection{\label{sec:d3}Diagram 3}
%%%%%%%%%%%%%%%%%%%%%%%%%%%%%%%%%%%%%%%%%%%%%%%%%%%%%%%%%%%%%%%%%%%%%%%%%%%%%%%%%%%%%%

\begin{figure}
    \includegraphics[width=.4\textwidth]{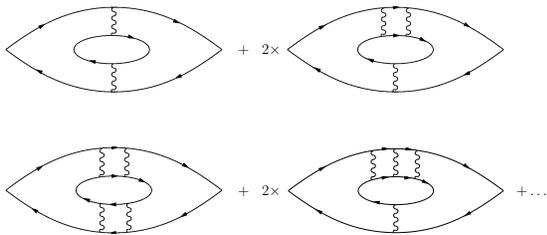}
    \caption{\label{fig:5}The series of diagrams contributing to $\delta\chi_{3}(Q)$. At the top, the second- and third-order order diagrams. Two equivalent third-order diagrams arise from the addition of a parallel interaction line to either the upper or the lower part of the second-order diagram. At the bottom, three fourth-order diagrams.}
\end{figure}

The last diagram $\delta\chi_{3}^{(2)}$ differs from those already discussed in the sense that it allows for the separate renormalization of either the upper or lower interaction line. This results in the appearance of two equivalent third-order diagrams and three fourth-order diagrams (of which two are equal), and so forth. These lowest order diagrams are shown in Fig.~\ref{fig:5}. Accordingly, we define the quantities $\delta\chi_{3}^{(i,j)}$, where ladders of order $i$ and $j$ are inserted in place of the original interaction lines. In particular, $\delta\chi_{3}^{(n)}=\sum_{i,j}\delta\chi_{3}^{(i,j)}\delta_{n,i+j}$ and 
\begin{equation}
\delta\chi_{3}(Q)=\sum_{i,j=1}^{\infty}\delta\chi_{3}^{(i,j)}(Q).
\end{equation}

The second difference stems from the fact that a finite nonanalytic correction is obtained from the leading terms in the particle-particle ladders of order $(\frac{m}{2\pi}\ln\frac{v_{F}Q}{W})^{i-1}$ and $(\frac{m}{2\pi}\ln\frac{v_{F}Q}{W})^{j-1}$, respectively. In fact, extracting this leading term from Eq.~(\ref{eq:PiNres}) we obtain
\begin{equation}
\Pi^{(j)}(P_\mu,\theta)=\sum_n V_n^j e^{-i n\theta} \Big(\frac{m}{2\pi}\ln\frac{v_{F}Q}{W}\Big)^{j-1}+\ldots~,
\end{equation}
and by performing the sum over $j$ we get
\begin{equation}
\sum_{j=1}^{\infty}\Pi^{(j)}(P_\mu,\theta)=\Gamma(\theta)+\ldots~.
\end{equation}
A similar argument can be repeated for the $i$th order interaction ladder. Therefore, the bare potential is replaced by renormalized expression (\ref{eq:Gamma_theta_def}) and the final result, 
\begin{equation}
    \delta\chi_{3}(Q)=K(0,Q)[\Gamma^{2}(\pi)-\Gamma^{2}(0)],
\end{equation}
is immediately obtained from Eq.~(\ref{eq:chi23}).

\subsection{Renormalized nonanalytic correction}

Combining the results of Sec.~\ref{sec:d124} and \ref{sec:d3}, it is clear that the final result has the same form of Eq.~(\ref{eq:2nd_order_final}) if $V(2k_F)$ is substituted by $\Gamma(\pi)$. The explicit expression reads as
\begin{equation}\label{eq:final_result_explicit}
    \delta\chi_s(Q)=\frac{m^3}{24\pi^4}\frac{Q}{k_F}\left[\sum_n \frac{V_n (-1)^n}{1-\frac{m V_n}{2\pi}\ln\frac{v_F Q}{W}}\right]^2.
\end{equation}

%%%%%%%%%%%%%%%%%%%%%%%%%%%%%%%%%%%%%%%%%%%%%%%%%%%%%%%%%%%%%%%%%%%%%%%%%%%%%%%%%%%%
\section{\label{sec:RG} Relation to the RG treatment}
%%%%%%%%%%%%%%%%%%%%%%%%%%%%%%%%%%%%%%%%%%%%%%%%%%%%%%%%%%%%%%%%%%%%%%%%%%%%%%%%%%%%

As discussed, our calculation was partially motivated by the renormalization group (RG) argument of Ref.~\onlinecite{PhysRevB.77.045108}. In this section, we further substantiate this argument.
Starting from  Eqs.~(\ref{eq:ppprop}) and (\ref{eq:Pi0Rphi}), one can calculate the second-order correction to the bare vertex $\Pi^{(1)}=\sum_n V_n e^{i n \theta }$ given by
\begin{equation}\label{eq:rg_second_order}
\Pi^{(2)}(P_\mu, \theta)=\frac{m}{2\pi}\ln \frac{v_FQ}{W} \sum_n V_n^2 e^{i n \theta} + \ldots~, 
\end{equation}
where we explicitly extracted the dependence on the upper cutoff $W$. 
From Eq.~(\ref{eq:rg_second_order}), we can immediately derive the following RG equations for the scale-dependent couplings $\Gamma_n(\Lambda=v_FQ)$:
\begin{equation}\label{eq:RG}
\frac{ {\rm d} \Gamma_n}{{\rm d}\ln \frac{\Lambda}{W}}=\frac{m}{2\pi} \Gamma_n^2,
\end{equation}
as in Ref.~\onlinecite{PhysRevB.77.045108}. 
This leads to the standard Cooper channel renormalization. A direct derivation of these scaling equations can be found
in Ref.~\onlinecite{shankar_review}. At this lowest order, we obtain an infinite number of independent flow equations, one for each
angular momentum $n$. The integration of these scaling equations directly leads to Eq. (\ref{eq:GammaDef}). 
These flow equations tell us that the couplings $\Gamma_n$ are marginally relevant in the infrared limit 
when the bare $\Gamma_n$
are negative and marginally irrelevant otherwise. Notice that at zero temperature, the running flow parameter $\Lambda$ 
is replaced by the momentum $v_FQ$ in the Cooper channel.
The idea of the RG is to replace in the perturbative calculations of a momentum-dependent quantity the bare 
couplings $\Gamma_n$ by their
renormalized values. By doing so, we directly resum an infinite class of (ladder) diagrams. 

Let us apply this reasoning now to the susceptibility diagrams and note that the first nonzero
contribution to the linear-in-$Q$ behavior of $\chi_s(Q)$ appears in the second order in $\Gamma_n$.
For the particular example of $\delta\chi_3$, the renormalization procedure has to be carried out independently 
for the two interaction lines, as illustrated by the series of diagrams in Fig.~\ref{fig:5}. 
For a given order of the interaction ladder in the bottom (top) part of the diagram, 
one can perform the Cooper channel resummation of the top (bottom) interaction ladders to infinite order, 
as described in Sec.~\ref{sec:d3} or by using the RG equations. 
The fact that renormalized amplitudes $\Gamma_n$ appear in the final results for the remaining diagrams $\delta\chi_{1,2,4}$ is also clear from the RG argument, after insertion of particle-particle ladders as in Figs.~\ref{fig:3}--\ref{fig:6}. 

\begin{figure}
    \includegraphics[width=.4\textwidth]{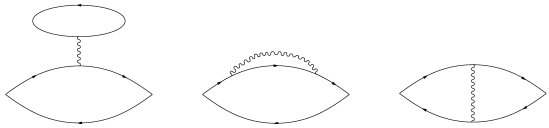}
    \caption{\label{fig:FirstOrder} First-order diagrams contributing to the spin susceptibility. These are renormalized by the leading logarithmic terms of the higher order diagrams (see Figs.~\ref{fig:3}--\ref{fig:6}). However, they do not produce a nonanalytic correction and can be neglected in the limit of small $Q$.}
\end{figure}

Finally, we note that the same series of diagrams that renormalizes the nonanalytic second-order contributions $\delta\chi^{(2)}_{1,2,4}$ also contributes to the renormalization of the first-order diagrams displayed in Fig.~\ref{fig:FirstOrder} (notice that the first one is actually vanishing because of charge neutrality). 
As it is clear from the explicit calculation Sec.~\ref{sec:hod}, the highest logarithmic powers, i.e., $\propto (\ln v_FQ/W)^{n-1}$ at order $n$, renormalize $V_m$ to $\Gamma_m$ in the final expressions for Fig.~\ref{fig:FirstOrder}. These first-order diagrams have an analytic dependence, at most $Q^2$. Therefore, in agreement with the discussion in Sec.~\ref{sec:d124}, the largest powers of the logarithms are not important for the linear dependence in $Q$ and, in fact, they were already neglected to second order.\cite{PhysRevB.68.155113}

%%%%%%%%%%%%%%%%%%%%%%%%%%%%%%%%%%%%%%%%%%%%%%%%%%%%%%%%%%%%%%%%%%%%%%%%%%%%%%%%%%%%
\section{\label{sec:con}Conclusions}
%%%%%%%%%%%%%%%%%%%%%%%%%%%%%%%%%%%%%%%%%%%%%%%%%%%%%%%%%%%%%%%%%%%%%%%%%%%%%%%%%%%%

In this paper we discussed the renormalization effects in the Cooper channel on the momentum-dependent spin susceptibility. The main result of the paper is given by Eq.~(\ref{eq:final_result_explicit}) and shows that each harmonics gets renormalized independently. The derivation of the higher order corrections to the spin susceptibility was based on the second-order result, which we revisited through an independent direct calculation in the particle-particle channel. Taking the angular dependence of the scattering potential explicitly into account, we verified that the main contribution indeed enters through forward- and back-scattering processes. At higher order, we found a simple and efficient way of resumming all the diagrams which contribute to the Cooper renormalization. We identified the leading nonvanishing logarithm in each ladder and used this result in the second-order correction. This method saves a lot of effort and, in fact, makes the calculation possible.

It was argued elsewhere that these renormalization effects might underpin the nonmonotonic behavior of the electron spin susceptibility if the higher negative harmonics override the initially leading positive Fourier components. This would results in the negative slope of the spin susceptibility at small momenta or temperatures.\cite{PhysRevB.74.205122,PhysRevB.77.045108} Other effects neglected here, as subleading logarithmic terms and nonperturbative contributions beyond the Cooper channel renormalization,\cite{ProcNatlAcadSci.103.15765} become relevant far away from the Kohn-Luttinger instability condition, but a systematic treatment in this regime is outside the scope of this work. Our results could be also extended to include material-related issues such as disorder and spin-orbit coupling, which are possibly relevant in actual samples.

We also notice that final expression (\ref{eq:final_result_explicit}) parallels the temperature dependence discussed in Ref.~\onlinecite{PhysRevB.74.205122},
suggesting that the temperature and momentum dependence are qualitatively similar in two dimensions. This was already observed from the second-order calculation, in which a linear dependence both in $Q$ and $T$ is obtained. In our work we find that this correspondence continues to hold in the nonperturbative regime if the Cooper channel contributions are included. This conclusion is nontrivial and, in fact, does not hold for the three-dimensional case. 

The last remark, together with the experimental observation of Ref.~\onlinecite{PhysRevB.67.205407}, supports the recent prediction that the ferromagnetic ordering of nuclear spins embedded in the two-dimensional electron gas is possible.\cite{PhysRevLett.98.156401,PhysRevB.77.045108} The ferromagnetic phase would be stabilized by the long-range Ruderman-Kittel-Kasuya-Yosida (RKKY) interaction, as determined by the nonanalytic corrections discussed here. 

%%%%%%%%%%%%%%%%%%%%%%%%%%%%%%%%%%%%%%%%%%%%%%%%%%%%%%%%%%%%%%%%%%%%%%%%%%%%%%%%%%%%
\begin{acknowledgments}
We thank M.~Borhani and D.~Maslov for their insightful comments. We also acknowledge discussions with B.~Braunecker, O.~Chalaev, J. C. Egues, D. Hirashima, C. P\'epin, and G. Schwiete. This work was supported by the Swiss NSF and the NCCR Nanoscience Basel.
\end{acknowledgments}
%%%%%%%%%%%%%%%%%%%%%%%%%%%%%%%%%%%%%%%%%%%%%%%%%%%%%%%%%%%%%%%%%%%%%%%%%%%%%%%%%%%%

\appendix

\begin{widetext}

%%%%%%%%%%%%%%%%%%%%%%%%%%%%%%%%%%%%%%%%%%%%%%%%%%%%%%%%%%%%%%%%%%%%%%%%%%%%%%%%%%%%
\section{\label{app:lad} DERIVATION OF THE LADDER DIAGRAMS}
%%%%%%%%%%%%%%%%%%%%%%%%%%%%%%%%%%%%%%%%%%%%%%%%%%%%%%%%%%%%%%%%%%%%%%%%%%%%%%%%%%%%

To calculate the ladder diagram given by Eq.~(\ref{eq:Pidef}) we begin from the simultaneous change of all $q_{i}$ variables, $q_{i,\mu}\rightarrow-q_{i,\mu}-p_{\mu}$, and expand the scattering potential into
its Fourier components given by Eq.~(\ref{eq:Vdef}). Thus,
\begin{equation}\label{eq:ppappendix_1}
    \Pi^{(n)}(p_\mu,p'_\mu,k_\mu) = \sum_{m_{1}\dots m_{n}} (-1)^{n-1+m_1-m_n}
    V_{m_{1}}\dots V_{m_{n}}e^{-im_{1}\theta_{p}+im_{n}\theta_{p'}}\prod_{i=1}^{n-1}\int\frac{{ d}^{3}q_{i}}
    {(2\pi)^{3}}G(-q_{i,\mu})G(q_{i,\mu}+P_{\mu})
    e^{i(m_{i}-m_{i+1})\theta_{q_{i}}},
\end{equation}
where $P_{\mu}\equiv k_{\mu}+p_{\mu}$. We first evaluate the factors
$\Pi_{M}(P_\mu) \equiv -\int\frac{{ d}^{3}q}{(2\pi)^{3}}G(-q_{\mu})G(q_{\mu}+P_{\mu})e^{iM\theta_{q}}$ appearing in the above formula. To this end, we integrate over the frequency $\Omega_{q}$ and linearize the spectrum around the Fermi surface, $\xi_{\bm{q}+\bm{P}}\approx\xi_{\bf q}+v_{F}P\cos\theta_{q}$. This requires that $\theta_q$, and all the angles in Eq.~(\ref{eq:ppappendix_1}), are defined from the direction of ${\bf P}$. We also use $\xi=\xi_{\bf q}$ as integration variable, which gives
\begin{equation}
    \Pi_{M}(P_\mu) = -\frac{m}{(2\pi)^{2}}\int_0^{2\pi} { d}\theta_{q} e^{iM\theta_{q}} \int { d}\xi \,
    \frac{\Theta(\xi+v_{F}P\cos\theta_{q})-\Theta(-\xi)}
    {2\xi+v_{F}P\cos\theta_{q}-i\Omega_{P}}.
\end{equation}
The unit step functions, $\Theta(\xi+v_{F}P\cos\theta_{q})$ and $\Theta(-\xi)$, determine the integration range in $\xi$, which is $-v_{F}P\cos\theta_{q}$ to $W$ and $-W$ to $0$, respectively, with $W$ being the high energy cutoff. The energy integration yields
\begin{equation}
    \Pi_{M}(P_\mu) = -\frac{m}{(2\pi)^{2}}\int_{0}^{2\pi}
    {d}\theta_{q} ~ e^{iM\theta}\, \frac12 \bigg(-i\pi\tm{sgn}\Omega_{P}
    +\ln\frac{2W}{-v_{F}P\cos\theta_{q}-i\Omega_{P}}
    +\ln\frac{2W}{v_{F}P\cos\theta_{q}-i\Omega_{P}}\bigg),
\end{equation}
with the sign term coming from the lower limit of the second integration, $\ln(-2W-i\Omega_{P})=\ln2W-i\pi\tm{sgn}\Omega_{P}$.

If we change variables in the integral over the first logarithm, $\theta_{q}\rightarrow\theta_{q}+\pi$, we make it identical to the second one, except for the multiplicative term $(-1)^{M}$ originating from $e^{iM(\theta_{q}+\pi)}$. Therefore, we find that
\begin{equation}
    \Pi_{M}(P_\mu) = \frac{m}{(2\pi)^{2}}\int_{0}^{2\pi}
    { d}\theta_{q} ~ e^{iM\theta_{q}}\bigg[\frac{i\pi}{2}\tm{sgn}\Omega_{P}+\ln\frac{-i\Omega_{P}}{W}
    +\ln\Big(1+i\frac{v_{F}P}{\Omega_{P}}\cos\theta_{q}\Big)\bigg],
\end{equation}
where $M$ is even and the factor of $2$ in front of $W$ has been absorbed into the cutoff. Writing the second logarithm as a series, $\ln(1+x)=-\sum_{n=1}^{\infty}(-1)^{n}\frac{x^{n}}{n}$, we can easily integrate term by term. 

The $M=0$ contribution is
\begin{equation}
    \Pi_{0} (P_\mu)=
    \frac{m}{(2\pi)^{2}}\int_{0}^{2\pi}{d}\theta_{q}\bigg[\ln\frac{|\Omega_{P}|}{W}
    -\sum_{n=1}^{\infty}\frac{1}{n}\Big(\frac{-iv_{F}P}{\Omega_{P}}\cos\theta_{q}\Big)^{n}\bigg]=
    \frac{m}{2\pi}\bigg[\ln\frac{|\Omega_{P}|}{W} - {\sum_{n\geq2}}^{\prime}
    \frac{1}{n}\binom{n}{\frac{n}{2}}\Big(\frac{-i v_{F}P}{2\Omega_{P}}\Big)^{n}\bigg],
\end{equation}
where $n$ is even in the primed sum. The summation of the series gives Eq.~(\ref{eq:Pi0}) shown in the main text.

For $M\neq0$ ($M$ even) we get
\begin{equation}
    \Pi_{M}(P_\mu) = -\frac{m}{(2\pi)^{2}}\int_{0}^{2\pi}{d}\theta_{q}
    e^{iM\theta_{q}}\sum_{n}\frac{1}{n}\Big(\frac{-i v_{F}P}{\Omega_{P}}\cos\theta_{q}\Big)^{n}= -\frac{m}{2\pi}{\sum_{n\geq|M|}}^{\hspace{-5pt}\prime}~
    \frac{1}{n}\binom{n}{\frac{n-|M|}{2}}\Big(\frac{-i v_{F}P}{2\Omega_{P}}\Big)^{n}.
\end{equation}
Upon summation over even values of $n$, Eq.~(\ref{eq:PiM}) in the main text is obtained.
We now consider Eq.~(\ref{eq:ppappendix_1}), and to simplify the notation we introduce summation indexes $M_{n-i}\equiv m_{i}-m_{i+1}$ (for $i=1,2,\ldots n-1$) and $m\equiv -m_n$. We also introduce $m'=m_1-m_n=\sum_i M_i$, which is even. We express the angles from the direction of ${\bf p}$, i.e., $\theta_{p}\to -\theta_P$ and $\theta_{p'}\to \theta-\theta_P$. Finally, we obtain
\begin{equation}
    \Pi^{(n)}(P_\mu,\theta) = {\sum_{m',M_{1}\dots M_{n-1}}}^{\hspace{-19pt}\prime}~~~~~
    V_{-m}V_{M_1-m}\dots V_{M_1+\ldots M_{n-1}-m}e^{i m'\theta_{P}-i m\theta}\Pi_{M_1}(P_\mu)\ldots \Pi_{M_{n-1}}(P_\mu),
\end{equation}
which can be written as in Eq.~(\ref{eq:PiNres}) using definition (\ref{eq:VMNdef}). The special case of $\Pi^{(2)}$ is given by Eq.~(\ref{eq:ppprop}) and can be directly used for the derivation of RG equations in Sec.~\ref{sec:RG}.

%%%%%%%%%%%%%%%%%%%%%%%%%%%%%%%%%%%%%%%%%%%%%%%%%%%%%%%%%%%%%%%%%%%%%%%%%%%%%%%%%%%%
\section{\label{app:chi1-2} GREEN'S FUNCTIONS INTEGRATION FOR $\delta\chi_{1}^{(n)}$}
%%%%%%%%%%%%%%%%%%%%%%%%%%%%%%%%%%%%%%%%%%%%%%%%%%%%%%%%%%%%%%%%%%%%%%%%%%%%%%%%%%%%

We consider here the integration of the Green's functions appearing explicitly in Eqs.~(\ref{eq:chi1-2def}) and (\ref{eq:chi1-ndef});
\begin{align}\label{eq:GFintegral}
    A\equiv 
    -\int\frac{{d}^{2}k}{(2\pi)^{2}}\lim_{\xi_{\bm{k}'}\rightarrow\xi_{\bf k}}
    \int\frac{{d}\omega}{2\pi}~\frac{1}{(\omega+i\xi_{\bf k})(\omega+i\xi_{\bm{k}'})
    (\omega+i\xi_{\bm{k}+\bm{Q}})
    (\omega-\Omega_{P}-i\xi_{\bm{k}-\bm{P}})}.
\end{align}
The integration over $\omega$ can be performed with the method of residues (we choose the lower half-plane contour); 
\begin{align}\label{eq:GFs_integrated1}
    \notag & A = im\int\frac{{ d}\theta_k { d}\xi_{\bm{k}}}{(2\pi)^{2}}
    \lim_{\xi_{\bm{k}'}\rightarrow\xi_{\bm{k}}}
    \bigg[\frac{\Theta(\xi_{\bm{k}})}{(\xi_{\bm{k}'}-\xi_{\bm{k}})
    (\xi_{\bm{k}+\bm{Q}}-\xi_{\bm{k}})
    (\Omega_{P}+i\xi_{\bm{k}}+i\xi_{\bm{k}-\bm{P}})}
    +\frac{\Theta(\xi_{\bm{k}'})}{(\xi_{\bm{k}}-\xi_{\bm{k}'})
    (\xi_{\bm{k}+\bm{Q}}-\xi_{\bm{k}'})
    (\Omega_{P}+i\xi_{\bm{k}'}+i\xi_{\bm{k}-\bm{P}})}\\
    &+\frac{\Theta(\xi_{\bm{k}+\bm{Q}})}{(\xi_{\bm{k}}-\xi_{\bm{k}+\bm{Q}})
    (\xi_{\bm{k'}}-\xi_{\bm{k}+\bm{Q}})
    (\Omega_{P}+i\xi_{\bm{k}+\bm{Q}}+i\xi_{\bm{k}-\bm{P}})}
    +\frac{\Theta(-\xi_{\bm{k}-\bm{P}})}
    {(\Omega_{P}+i\xi_{\bm{k}}+i\xi_{\bm{k}-\bm{P}})
    (\Omega_{P}+i\xi_{\bm{k}'}+i\xi_{\bm{k}-\bm{P}})
    (\Omega_{P}+i\xi_{\bm{k}+\bm{Q}}+i\xi_{\bm{k}-\bm{P}})}\bigg].
\end{align}

Since the sum of all residues of the integrand in Eq.~(\ref{eq:GFintegral}) in the entire complex plane is zero, we can subtract the same quantity $\Theta(\xi_{\bm{k}})$ from each numerator in Eq.~(\ref{eq:GFs_integrated1}) without affecting the result. This cancels the first term and, using $\lim_{\xi_{\bm{k}'}\rightarrow\xi_{\bm{k}}}\frac{{\Theta(\xi_{\bm{k}'})-\Theta(\xi_{\bm{k}})}} {\xi_{\bm{k}'}-\xi_{\bm{k}}} = \delta(\xi_{\bm{k}})$,  we obtain 
\begin{align}
    \notag A = A_{1}+A_{2}+A_{3}
    =&im\int\frac{{ d}\theta_k { d}\xi_{\bm{k}}}{(2\pi)^{2}}\bigg[
    \frac{i\delta(\xi_{\bm{k}})}{(\xi_{\bm{k}}-\xi_{\bm{k}+\bm{Q}})
    (\Omega_{P}+i\xi_{\bm{k}}+i\xi_{\bm{k}-\bm{P}})} \\
    &+\frac{\Theta(\xi_{\bm{k}+\bm{Q}})-\Theta(\xi_{\bm{k}})}
    {(\xi_{\bm{k}}-\xi_{\bm{k}+\bm{Q}})^{2}
    (\Omega_{P}+i\xi_{\bm{k}+\bm{Q}}+i\xi_{\bm{k}-\bm{P}})}
    +\frac{\Theta(-\xi_{\bm{k}-\bm{P}})-\Theta(\xi_{\bm{k}})}
    {(\Omega_{P}+i\xi_{\bm{k}}+i\xi_{\bm{k}-\bm{P}})^{2}
    (\Omega_{P}+i\xi_{\bm{k}+\bm{Q}}+i\xi_{\bm{k}-\bm{P}})}\bigg].
\end{align}

We can now perform the integration in $d\xi_{\bm{k}}$. To this end we linearize the energy spectrum near the Fermi surface, $\xi_{\bm{k}+\bm{Q}}\approx\xi_{\bf k}+v_{F}Q\cos\theta_k$ and
$\xi_{\bm{k}-\bm{P}}\approx\xi_{\bf k}-v_{F}P\cos\theta_P$, which is a good approximation since $Q,P\ll k_F$. We also define $\zeta\equiv\Omega_{P}-iv_{F}P\cos\theta_P$ (and the complex conjugate $\bar\zeta\equiv\Omega_{P}+iv_{F}P\cos\theta_P$). The first two integrals are vanishing, 
\begin{equation}
    \notag A_{1} = \frac{m}{(2\pi)^{2}}\int_{0}^{2\pi}{ d}\theta_k
    \frac{1}{\zeta v_{F}Q\cos\theta_k}=0, \qquad {\rm and} \qquad
    \notag A_{2} = \frac{m}{8\pi^{2}}\int_{0}^{2\pi}\frac{{ d}\theta_k}{(v_{F}Q\cos\theta_k)^{2}}
    \ln\bigg(\frac{\zeta+iv_{F}Q\cos\theta_k}{\zeta-iv_{F}Q\cos\theta_k}\bigg)=0,
\end{equation}
since the integrands are odd with respect to $\cos\theta_k$. 

For the remaining term we make use of the indefinite integral,
\begin{equation}
    \notag \int\frac{{ d}\xi}{(2i\xi+z_{1})^{2}(2i\xi+z_{2})}
    =\frac{z_{1}-z_{2}+(z_{1}+2i\xi)\ln\big(\frac{z_{2}+2i\xi}
    {z_{1}+2i\xi}\big)}{2i(z_{1}-z_{2})^{2}(2i\xi+z_{1})},
\end{equation}
which tends to zero for $\xi\rightarrow\pm\infty$, and
hence only the integration limits at $0$ and $v_{F}P\cos\theta_{P}$
contribute;
\begin{equation}
    A = \frac{m}{8\pi^{2}}\int_{0}^{2\pi}\bigg\{
    \frac{1}{(v_{F}Q\cos\theta_k)^{2}}
    \bigg[\ln\bigg(\frac{\bar{\zeta}}{\bar{\zeta}+iv_{F}Q\cos\theta_k}\bigg)
    +\ln\bigg(\frac{\zeta}{\zeta+iv_{F}Q\cos\theta_k}\bigg)\bigg]
    +\frac{2i\Omega_P}{|\zeta|^{2}v_{F}Q\cos\theta_k}\bigg\}{d}\theta_k.
\end{equation}
The last term vanishes and the final integration in $d\theta_k$ can be done using
\begin{equation}
    \int_{0}^{2\pi}~\frac{\ln\big(\frac{z}{z + ia\cos\theta}\big)}{a^2\cos^2\theta}~{ d}\theta
    =\frac{2\pi}{a^{2}}\Big(1-\tm{sgn}(\tm{Re}z)\frac{\sqrt{z^{2}+a^{2}}}{z}\Big),
\end{equation}
which yields
\begin{equation}
    A = \frac{m}{4\pi(v_{F}Q)^2}\Bigg[ 2-
    \frac{\sqrt{(\Omega_{P}+iv_{F}P\cos\theta_P)^{2}+(v_{F}Q)^2}}{\Omega_{P}+iv_{F}P\cos\theta_P}\tm{sgn}\,\Omega_P   -\frac{\sqrt{(\Omega_{P}-iv_{F}P\cos\theta_P)^{2}+(v_{F}Q)^2}}{\Omega_{P}-iv_{F}P\cos\theta_P}\tm{sgn}\,\Omega_P\Bigg].
\end{equation}
Notice that $A$ is even in $\Omega_P$. Furthermore, the last two terms give the same contribution upon integration $\int_0^{2\pi}{ d}\theta_P$, and therefore Eq.~(\ref{eq:2ndorder_dimensional}) is obtained. 

%%%%%%%%%%%%%%%%%%%%%%%%%%%%%%%%%%%%%%%%%%%%%%%%%%%%%%%%%%%%%%%%%%%%%%%%%%%%%%%%%%%%
\section{\label{app:chi1_rederived}CALCULATION OF $\delta\chi_1^{(2)}$}
%%%%%%%%%%%%%%%%%%%%%%%%%%%%%%%%%%%%%%%%%%%%%%%%%%%%%%%%%%%%%%%%%%%%%%%%%%%%%%%%%%%%

In this appendix we consider the explicit evaluation of Eq.~(\ref{eq:chi1-2rescaled}). It receives contributions from all possible values of $M$ appearing in $\Pi^{(2)}$ [see Eq.~(\ref{eq:ppprop})]. We consider first $M=0$ for which Eq.~(\ref{eq:Pi0Rphi}) is useful. With the change of variables $r=R(\sin\phi+i\cos\phi\cos\theta_P)$ we obtain 
\begin{equation}\label{eq:chi1int}
    \delta\chi_{1,0}^{(2)}
    = -\frac{m^{2} Q}{2\pi^{5}v_{F}}\tilde{V}_0^{2}\int_{0}^{\pi/2}{ d}\phi\cos\phi
    \int_{0}^{2\pi}{d}\theta_P
    \int_{0}^{r_{max}(\phi,\theta_{P})}{ d}rr^{2}\Big(1-\frac{\sqrt{r^{2}+1}}{r}\Big)
    \frac{\ln\big(\frac{1+\sin\phi}{\sin\phi+i\cos\phi\cos\theta_P}\big)
    +\ln\big(\frac{v_F Qr}{W}\big)}{(\sin\phi+i\cos\phi\cos\theta_P)^{3}},
\end{equation}
where $r_{max}=\frac{W}{v_{F}Q}(\sin\phi+i\cos\phi\cos\theta_P)$. Note, that it is necessary to introduce the upper cutoff in the integral over $r$, which is formally divergent. However, this upper limit turns out to be irrelevant for the nonanalytic correction.

We start from the first contribution in the above equation, where the $r$ integration gives
\begin{equation}
    Q\int_{0}^{r_{max}}{d}rr^{2}\Big(1-\frac{\sqrt{r^{2}+1}}{r}\Big)
    = \frac{Q}{3}[r_{max}^{3}-(1+r_{max}^{2})^{3/2}+1]
    \approx \frac{W}{v_{F}}(\sin\phi+i\cos\phi\cos\theta_P)+\frac{Q}{3}.
\end{equation}
The term proportional to $W$, as in Ref.~\onlinecite{PhysRevB.68.155113}, is the dominant contribution to the spin susceptibility. However, it does not depend on $Q$ and therefore is uninteresting for us. The term proportional to $Q$, from the lower integration limit, is the desired nonanalytic correction to the spin susceptibility and does not depend on $\phi$ and $\theta_{P}$. Therefore, the angular integration can be performed using\cite{comment:integral}
\begin{equation}
    \int_{0}^{\pi/2}{ d}\phi\cos\phi\int_{0}^{2\pi}{ d}\theta_P
    \frac{\ln\big(\frac{1+\sin\phi}{\sin\phi+i\cos\phi\cos\theta_P}\big)}
    {(\sin\phi+i\cos\phi\cos\theta_P)^{3}} = -\frac{\pi}{2}.
\end{equation}
The same analysis can be applied to the second term of Eq.~(\ref{eq:chi1int}), which contains $\ln(v_{F}Qr/W)$. The integration in $r$ gives a constant from the lower limit and the remaining angular integrations yield zero, as discussed in the main text.

Hence, the final result is
\begin{equation}
    \delta\chi_{1,0}^{(2)}(Q) = \frac{m^{2}}{12\pi^{4}v_{F}}Q\sum_n V_n^{2}.
\end{equation}

We now aim to calculate terms with $M\neq0$. By making use of Eq.~(\ref{eq:PiM}) for $\Pi_M(R,\phi)$ and substituting again $r\equiv R(\sin\phi+i\cos\phi\cos\theta_P)$ we obtain
\begin{equation}
    \delta\chi_{1,M}^{(2)}
    =\frac{m^{2}Q}{2|M|\pi^{5}v_{F}}\sum_{n}V_{M-n}V_{n}\int_{0}^{\pi/2}{ d}\phi\int_{0}^{2\pi}{d}\theta_P
    \int_{0}^{r_{max}}{d}rr^{2}\Big(1-\frac{\sqrt{r^{2}+1}}{r}\Big)
    \Big(\frac{1-\sin\phi}{i\cos\phi}\Big)^{|M|}
    \frac{\cos\phi ~ e^{-iM\theta_P}}{(\sin\phi+i\cos\phi\cos\theta_P)^{3}}.
\end{equation}
The integral over $r$ can be performed as before. The integration over $\theta_P$ yields
\begin{equation}\label{eq:intphitheta}
    \int_{0}^{2\pi}{ d}\theta_P\frac{e^{-iM\theta_P}}
    {(\sin\phi+i\cos\phi\cos\theta_P)^{3}}=\pi(M^{2}+3|M|\sin\phi+3\sin^{2}\phi-1)
    \Big(\frac{1-\sin\phi}{i\cos\phi}\Big)^{|M|},
\end{equation}
which can be obtained by standard contour integration in the complex
plane ($z=e^{-i\theta_P}$). Finally,
\begin{equation}
    \delta\chi_{1,M}^{(2)} = \frac{m^{2}Q}{6|M|\pi^{4}v_{F}}\sum_{n}V_{M-n}V_{n}
    \int_{0}^{\pi/2}{d}\phi\cos\phi
    \Big(\frac{1-\sin\phi}{\cos\phi}\Big)^{2|M|}(M^{2}+3|M|\sin\phi+3\sin^{2}\phi-1)\\
    =\frac{m^{2}Q}{12\pi^{4}v_F}\sum_{n}V_{M-n}V_{n},
\end{equation}
since the last integration gives a factor of $|M|/2$. Thus, the total second order correction is
\begin{align}
    \delta\chi_{1}^{(2)}(Q) = \frac{m^{2}}{24\pi^{4}}
    \frac{Q}{v_{F}}{\sum_{M}}^{\prime}\sum_{n} 2 V_{M-n}V_{n}.
\end{align}
Rewriting the double sum as
$\sum_{m,n}V_{m}V_{n}+\sum_{m,n}(-1)^{m+n}V_{m}V_{n}$, we recover the
two contributions as in Eq.~(\ref{eq:chi21}).

%%%%%%%%%%%%%%%%%%%%%%%%%%%%%%%%%%%%%%%%%%%%%%%%%%%%%%%%%%%%%%%%%%%%%%%%%%%%%%%%%%%%%%%%%%%%%%%%%%%
\section{\label{app:ppsmallQ} Small $Q$ limit of $\Pi^{(n)}$}
%%%%%%%%%%%%%%%%%%%%%%%%%%%%%%%%%%%%%%%%%%%%%%%%%%%%%%%%%%%%%%%%%%%%%%%%%%%%%%%%%%%%%%%%%%%%%%%%%%%

We expand $\Pi^{(n)}$ [see Eq.~(\ref{eq:PiNres})] in powers of $\Pi_{0}(R,\phi)$ as follows
\begin{equation}\label{eq:PiNexp}
	 \Pi^{(n)}  (R,\phi,\theta_P,\theta)  \approx \tilde{V}^{n}_{00\dots0}  \Pi_{0}^{n-1}
	 + {\sum_{M\neq0}}^{\prime}\left(\tilde{V}^{n}_{M0\dots0}+ 
	\ldots \tilde{V}^{n}_{00\dots M}\right) \Pi_{M} \Pi_{0}^{n-2}+\ldots~.
\end{equation}
The expression of $\Pi_0(R,\phi)$ is given by Eq.~(\ref{eq:Pi0Rphi}). Therefore, the leading contribution in the above equation at small $Q$ is from the first term since $\Pi_{0}^{n-1}\approx (\frac{m}{2\pi}\ln\frac{v_{F}Q}{W})^{n-1}$. However, the leading order does not contribute to the nonanalytic correction. Neglecting such constant terms, we can write the relevant subleading contribution in the following way:
\begin{equation}	
	[\Pi_{0}(R,\phi)]^{n-1} = (n-1)\Big(\frac{m}{2\pi}\ln\frac{v_{F}Q}{W}\Big)^{n-2}
	\frac{m}{2\pi}\ln R(1+\sin\phi)+\ldots
	= (n-1)\Big(\frac{m}{2\pi}\ln\frac{v_{F}Q}{W}\Big)^{n-2}\Pi_0(R,\phi)+\ldots ~.
\end{equation}
Furthermore, by using $\Pi_{0}^{n-2}\approx (\frac{m}{2\pi}\ln\frac{v_{F}Q}{W})^{n-2}$, we can simplify Eq.~(\ref{eq:PiNexp}) to the following form:
\begin{equation}\label{eq:ppsmallQ_1}
	 \Pi^{(n)} (R,\phi,\theta_P,\theta)  = \Big(\frac{m}{2\pi}\ln\frac{v_{F}Q}{W}\Big)^{n-2} 
	 {\sum_{M}}^{\prime}\left(\tilde{V}^{n}_{M0\dots0}+\tilde{V}^{n}_{0M\dots0} 
	+\ldots \tilde{V}^{n}_{00\dots M}\right) \Pi_M + \ldots~.
\end{equation}
By making use of Eq.~(\ref{eq:VMNdef}) we have 
\begin{equation}
	\tilde{V}^{n}_{M0\dots0}+\tilde{V}^{n}_{0M\dots0} 
	+\ldots \tilde{V}^{n}_{00\dots M}
	 = \sum_{k}\sum_{j=1}^{n-1}V_{k}^{n-j}V_{k-M}^{j}
	e^{iM\theta_P-ik \theta}
\end{equation}
and therefore Eq.~(\ref{eq:ppsmallQ_1}) is written explicitly as
\begin{equation}\label{eq:piNexpansion}
	\Pi^{(n)} (R,\phi,\theta_P,\theta) = {\sum_M}^\prime \Pi_M(R,\phi)\sum_{k} V_{k}V_{k-M} e^{iM\theta_P-ik \theta}
	\sum_{j,j'= 0}^{\infty} \delta_{j+j',n-2}\left(\frac{m V_{k}}{2\pi}\ln\frac{v_{F}Q}{W} \right)^{j}
	\left(\frac{m V_{k-M}}{2\pi}\ln\frac{v_{F}Q}{W} \right)^{j'}+\ldots ~.
\end{equation}
We can now sum previous expression (\ref{eq:piNexpansion}) over the index $n\geq 2$, which removes the constraint $j+j'=n-2$. Hence, the last double summation factorizes in the product of two geometric series, that can be evaluated explicitly, and we obtain Eq.~(\ref{eq:finalPinSum}).

\end{widetext}


\begin{thebibliography}{50}

\bibitem{landau57}
L. D. Landau, Sov. Phys. JETP {\bf 3}, 920 (1957).

\bibitem{landau59}
L. D. Landau, Sov. Phys. JETP {\bf 8}, 70 (1959).

\bibitem{pines}
D.~Pines and P.~Nozi\'eres, \emph{The Theory of Quantum Liquids} (W. A. Benjamin, Inc., New York, 1966).

\bibitem{gfg}
G.~F. Giuliani and G. Vignale, \emph{Quantum Theory of the Electron Liquid} (Cambridge
University Press, Cambridge, 2005).

\bibitem{pethick73}
C. J. Pethick and G. M. Carneiro, Phys. Rev. A {\bf 7}, 304 (1973), and references therein.

\bibitem{coffey93}
D. Coffey and K. S. Bedell, Phys. Rev. Lett. {\bf 71}, 1043 (1993).

\bibitem{PhysRevB.55.9452}
D.~Belitz, T.~R.~Kirkpatrick, and T.~Vojta, Phys. Rev. B {\bf 55}, 9452 (1997).

\bibitem{PhysRevB.68.155113}
A.~V.~Chubukov and D.~L.~Maslov, Phys. Rev. B \textbf{68}, 155113 (2003).

\bibitem{greywall83}
D. S. Greywall, Phys. Rev. B {\bf 27}, 2747 (1983).

\bibitem{casey03}
A. Casey, H. Patel, J. Ny\'eki, B. P. Cowan, and J. Saunders, Phys. Rev. Lett. {\bf 90}, 115301 (2003).

\bibitem{PhysRevB.16.1933}
G.~M.~Carneiro and C.~J.~Pethick, Phys. Rev. B {\bf 16}, 1933 (1977).

\bibitem{hirashima98}
D. S. Hirashima and H. Takahashi, J. Phys. Soc. Jpn. {\bf 67}, 3816 (1998).

\bibitem{JETPLett.58.709}
M. A. Baranov, M. Yu. Kagam, and M. S. Mar'enko, JETP Lett. {\bf 58}, 709 (1993).

\bibitem{PhysRevLett.86.5337}
G. Y. Chitov and A. J. Millis, Phys. Rev. Lett. {\bf 86}, 5337 (2001)

\bibitem{PhysRevB.64.054414}
G. Y. Chitov and A. J. Millis, Phys. Rev. B {\bf 64}, 054414 (2001)

\bibitem{chubukov05a}
A. V. Chubukov, D. L. Maslov, S. Gangadharaiah, and L. I. Glazman,
Phys. Rev. Lett. {\bf 95}, 026402 (2005).

\bibitem{chubukov05b}
A. V. Chubukov, D. L. Maslov, S. Gangadharaiah, and L. I. Glazman,
Phys. Rev. B {\bf 71}, 205112 (2005).

\bibitem{chubukov06}
A. V. Chubukov, D. L. Maslov, and A. J. Millis, Phys. Rev. B {\bf 73}, 045128 (2006).

\bibitem{chubukov07}
A. V. Chubukov and D. L. Maslov, Phys. Rev. B {\bf 76}, 165111 (2007).

\bibitem{aleiner06}
I. L. Aleiner and K. B. Efetov, Phys. Rev. B {\bf 74}, 075102 (2006).

\bibitem{PhysRevB.74.205122}
A.~Shekhter and A.~M.~Finkel'stein, Phys. Rev. B \textbf{74}, 205122 (2006).

\bibitem{ProcNatlAcadSci.103.15765}
A.~Shekhter and A.~M.~Finkel'stein, Proc. Natl. Acad. Sci. U.S.A. \textbf{103}, 15765 (2006).

\bibitem{schwiete06}
G. Schwiete and K. B. Efetov, Phys. Rev. B {\bf 74}, 165108 (2006).

\bibitem{PhysRevB.67.205407}
O.~Prus, Y.~Yaish, M.~Reznikov, U.~Sivan, and V.~Pudalov, Phys. Rev. B \textbf{67}, 205407 (2003).

\bibitem{PhysRevLett.15.524}
W.~Kohn and J.~M.~Luttinger, Phys. Rev. Lett. \textbf{15}, 524 (1965).

\bibitem{PhysRevB.48.1097}
A.~V.~Chubukov, Phys. Rev. B \textbf{48}, 1097 (1993).

\bibitem{PhysRevB.77.045108}
P.~Simon, B.~Braunecker, and D.~Loss, Phys. Rev. B \textbf{77}, 045108 (2008).

\bibitem{PhysRevLett.98.156401}
P.~Simon and D.~Loss, Phys. Rev. Lett. \textbf{98}, 156401 (2007).

\bibitem{PhysRevB.67.144520}
V.~M.~Galitski and S.~Das~Sarma, Phys. Rev. B \textbf{67}, 144520 (2003).

\bibitem{PhysRevB.71.045338}
D.~S.~Saraga, B.~L.~Altshuler, D.~Loss, and R.~M.~Westervelt, Phys. Rev. B \textbf{71}, 045338 (2005).

\bibitem{Mahan}
G.~D.~Mahan, \textit{Many-Particle Physics} (Plenum Press, New York, 2000).

\bibitem{shankar_review}
R.~Shankar, Rev. Mod. Phys. \textbf{66}, 129 (1994).

\bibitem{comment:integral}
This result has been obtained by accurate numerical integration.

\end{thebibliography}
\end{document}